\documentclass[twocolumn,superscriptaddress,showpacs,floatfix,aps,prb]{revtex4}
\usepackage{graphicx}
\usepackage{dcolumn}
\usepackage{amsmath}
\usepackage{amssymb}
\usepackage{amsbsy}
\usepackage{textcomp}
\usepackage{units}
\usepackage{color}
\usepackage{units}
\usepackage[normalem]{ulem}
\usepackage{verbatim}
\usepackage{hyperref}
\usepackage{url}

\usepackage{multirow}

\usepackage{float}

\newcommand{\Rb}{RbFe$_2$As$_2$}
\newcommand{\Sr}{SrFe$_2$As$_2$}
\newcommand{\SR}{Sr$_{1-x}$Rb$_{x}$Fe$_2$As$_2$}

\newcommand{\Cs}{CsFe$_2$As$_2$}

\newcommand{\AeFeAs}{\textit{Ae}Fe$_2$As$_2$}
\newcommand{\AFeAs}{\textit{A}Fe$_2$As$_2$}

\begin{document}

\title{Evolution from $B_{2g}$ nematics to $B_{1g}$ nematics in heavily
hole-doped iron-based superconductors}

\author{Vladislav Borisov}

\affiliation{Institute of Theoretical Physics, Goethe University Frankfurt am
Main, D-60438 Frankfurt am Main, Germany}
\email[Corresponding author:\ ]{borisov@itp.uni-frankfurt.de}

\author{Rafael M. Fernandes}

\affiliation{School of Physics and Astronomy, University of Minnesota, Minneapolis,
Minnesota 55455, USA}

\author{Roser Valent\'i}

\affiliation{Institute of Theoretical Physics, Goethe University Frankfurt am
Main, D-60438 Frankfurt am Main, Germany}

\date{\today}
\begin{abstract}
Recent experiments reported an unusual nematic behavior of heavily
hole-doped pnictides \textit{A}Fe$_{2}$As$_{2}$\textit{\emph{, with
alkali }}\textit{A} = Rb, Cs. In contrast to the $B_{2g}$ nematic
order of the parent \textit{A}\emph{e}Fe$_{2}$As$_{2}$ compounds
(\textit{\emph{with alkaline earth }}\textit{Ae} = Sr, Ba), characterized
by unequal nearest-neighbor Fe-Fe bonds, in the hole-doped systems
nematic order is observed in the $B_{1g}$ channel, characterized
by unequal next-nearest-neighbor Fe-Fe (diagonal Fe-As-Fe) bonds.
In this work, using density functional theory, we attribute this behavior
to the evolution of the magnetic ground state along the series \emph{Ae}$_{1-x}$\emph{A}$_{x}$Fe$_{2}$As$_{2}$,
from single stripes for small $x$ to double stripes for large $x$.
Our simulations using the reduced Stoner theory show that fluctuations
of Fe moments are essential for the stability of the double-stripe
configuration. We propose that the change in the nature of the magnetic
ground state is responsible for the change in the symmetry of the
vestigial nematic order that it supports. 
\end{abstract}

\maketitle

\emph{Introduction.} Unconventional superconductivity in iron-based
superconductors is believed to be mediated by spin fluctuations and,
therefore, closely related to the nature of the magnetic interactions
\cite{Hirschfeld11,Chubukov11}. The most common type of magnetic
order~\textendash{} single-stripe phase~\textendash{} is observed
is many parent compounds, such as \AeFeAs{} (\textit{Ae} = Ca, Sr,
Ba) \cite{review1,review21,review3}. Stripe magnetic order is characterized
by an ordering vector $(\pi,0)$ or $(0,\pi)$ in the Fe-only square
lattice. As such, it breaks the tetragonal symmetry of the lattice
by making the nearest-neighbor Fe-Fe bonds inequivalent. External
pressure or chemical substitution suppress this magnetic order and
induce superconductivity (SC), suggesting a close link between the
two phases. Besides possibly providing the glue that binds Cooper
pairs together, magnetic fluctuations are also believed to be responsible
for the nematic phase of the \AeFeAs{} compounds \cite{Fernandes14}.
This is corroborated by the experimental observation that the magnetic
and nematic transition lines follow each other closely, and that the
elastic and magnetic fluctuations obey a scaling relationship \cite{Fernandes13,Bohmer16}.
One of the proposed mechanisms by which nematicity arises before the
onset of single-stripe magnetic order is that magnetic fluctuations
form a composite order parameter that condenses above the magnetic
transition. In this scenario, nematic order is a vestigial phase of
the magnetically ordered phase, breaking a subset of the symmetries
broken by the latter (i.e. tetragonal symmetry) while preserving other
symmetries (i.e. time-reversal symmetry) \cite{Sachdev08,Kivelson08,Fernandes12,Dagotto13}.

\begin{figure}[h]
\centering \vspace{-10pt}
 \includegraphics[width=0.95\columnwidth]{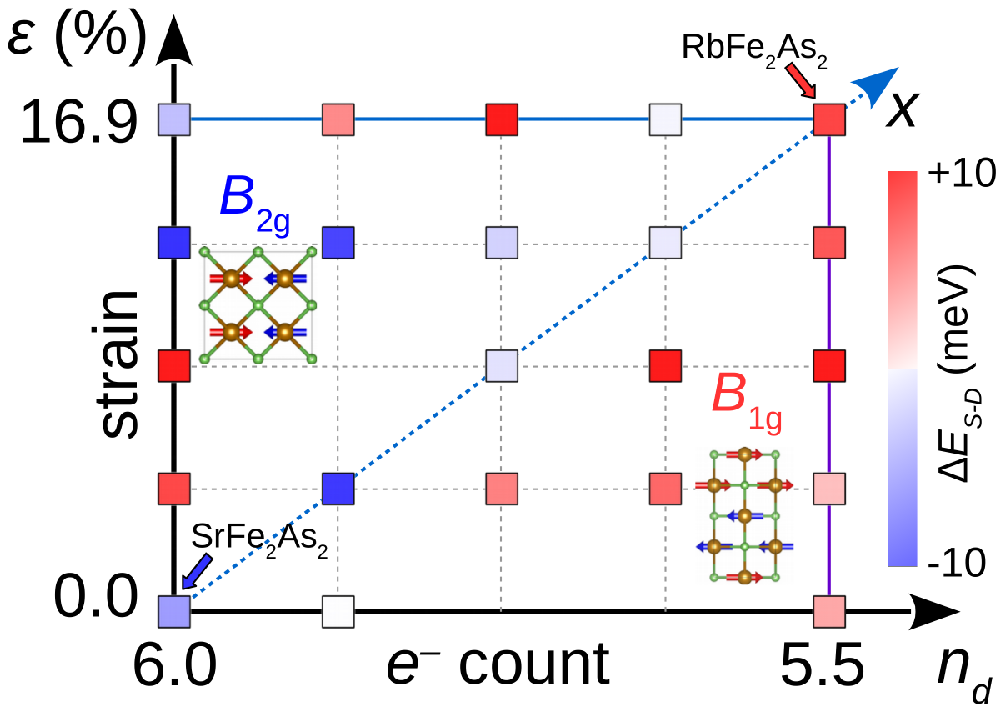} \vspace{-10pt}
 \caption{Schematic magnetic phase diagram of \SR{} as function of strain
(in $\%$) along the $c$-axis and Fe 3\textit{d} electron count.
Both parameters are tuned simultaneously as function of $x$, interpolating
between the two end compounds, \Sr{} ($\varepsilon=0\%$ and $n_{d}=6.0$)
and \Rb{} ($\varepsilon=16.9\%$ and $n_{d}=5.5$). The two types
of magnetic ground states are shown by color (red~\textendash{} double
stripe, blue~\textendash{} single stripe), and the color intensity
reflects the energy difference $\Delta E_{\mathrm{S-D}}=E_{\mathrm{single}}-E_{\mathrm{double}}$
between these two magnetic orders (in meV/Fe).}
\label{f:phase_diagram} 
\end{figure}

\begin{figure}
\begin{centering}
\includegraphics[width=0.99\columnwidth]{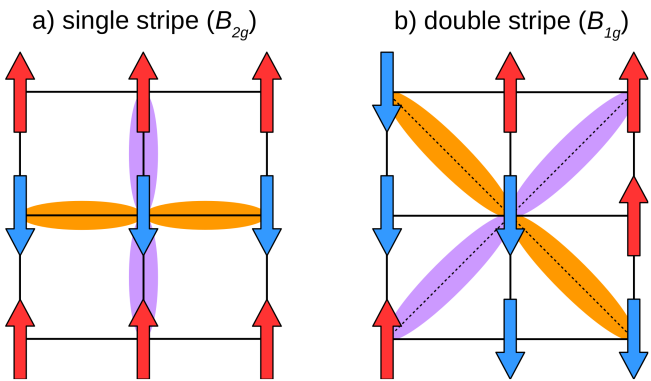} 
\par\end{centering}
\caption{Schematic representations of (a) the vestigial $B_{2g}$ nematic phase
associated with the single-stripe magnetic phase and (b) the vestigial
$B_{1g}$ nematic phase supported by the double-stripe magnetic phase.
The lobes illustrate the spin-spin correlation functions entering
the Ising-like composite order parameters $\Phi_{B_{2g}}=\left\langle \mathbf{S}_{\mathbf{R}_{i}}\cdot\mathbf{S}_{\mathbf{R}_{i}+\mathbf{x}}\right\rangle -\left\langle \mathbf{S}_{\mathbf{R}_{i}}\cdot\mathbf{S}_{\mathbf{R}_{i}+\mathbf{y}}\right\rangle $
and $\Phi_{B_{1g}}=\left\langle \mathbf{S}_{\mathbf{R}_{i}}\cdot\mathbf{S}_{\mathbf{R}_{i}+\mathbf{x}+\mathbf{y}}\right\rangle -\left\langle \mathbf{S}_{\mathbf{R}_{i}}\cdot\mathbf{S}_{\mathbf{R}_{i}-\mathbf{x}+\mathbf{y}}\right\rangle $.}
\label{fig_vestigial} 
\end{figure}

Very recently, a different type of nematic order was found in the
stoichiometric \Rb{} and \Cs{} compounds \cite{Li2016,Liu2018,Shibauchi18}.
Because these materials have an $n_{d}=5.5$ filling of the Fe $3d$
orbitals, in contrast to the $n_{d}=6$ filling of the \AeFeAs~
compounds, they are heavily hole-doped versions of the latter. The
nematic order displayed by \Rb{} lowers the tetragonal symmetry
by breaking the equivalence between the next-nearest-neighbor Fe-Fe
bonds, which correspond to the diagonals of the Fe-only square lattice.
In group-theory language, this corresponds to a $B_{1g}$ nematic
state, whereas the nematic state of \AeFeAs{} is in the $B_{2g}$
channel (see insets in Fig. \ref{f:phase_diagram} and \ref{fig_vestigial}).
Here, $B_{1g}$ and $B_{2g}$ refer to the irreducible representations
of the tetragonal point group associated with the actual crystallographic
unit cell, which contains two Fe atoms and is rotated $45^{\circ}$
with respect to the Fe-only square unit cell. An important question
that remains unaddressed is the origin of this $B_{1g}$ nematic state.
A recent theoretical study\cite{Onari2018} proposes that this new
type of nematicity is driven by incommensurate spin fluctuations on
$d_{xy}$ orbitals, and is related to the Lifshitz transition as the
system goes from $n_{d}=6$ to $n_{d}=5.5$ filling. On the other
hand, structural differences between these systems can, in principle,
produce significant changes in the magnetic interactions too, similarly
to the well-known case of FeSe$_{x}$Te$_{1-x}$.\cite{Moon2010}
Moreover, electronic correlations are believed to be much stronger
in \Rb{} and \Cs{} as compared to \AeFeAs, as manifested by the
larger Sommerfeld coefficients of the former \cite{Meingast16} and
suggested by theoretical studies based on the LDA+DMFT method.\cite{Yin2011,Medici2014,Diehl2014,Backes2015}

In this paper, we address the change of nematic order from \textbf{$B_{2g}$}
to $B_{1g}$ as a function of hole-doping in a specific 122 family
Sr$_{1-x}$Rb$_{x}$Fe$_{2}$As$_{2}$. In particular, we use first-principles
density functional theory (DFT) calculations to analyze the competition
between different magnetic orders as the iron 3\textit{d} orbital
filling evolves from $6$ to $5.5$ for increasing $x$. These DFT
calculations include the effects of i) hole doping, ii) structural
changes and, iii) Fe moment reduction by spin fluctuations within
the recently proposed reduced Stoner theory.\cite{Ortenzi2012,Ortenzi2015}
We find that the favored magnetic ground state changes from single-stripe
to double-stripe, which is characterized by an ordering vector of
$\left(\pi/2,\,\pi/2\right)$ in the Fe-only square Brillouin zone.
The main results are shown in the phase diagram of Fig. \ref{f:phase_diagram}.
The stabilization of the double-stripe order in \Rb, as compared
to the single-stripe order in \Sr, arises primarily from the changes
in the electron count, rather than the changes in the lattice structure,
and seems to be unrelated to Fermi surface nesting. Since the $B_{2g}$
nematic order is a vestigial phase of the single-stripe phase, whereas
the $B_{1g}$ nematic order is a vestigial phase of the double-stripe
phase, our results provide a unified picture in which the change in
the nematic state as function of hole-doping reflects a change in
the dominant magnetic interactions.

\emph{First-principle results: magnetic order.} We focus on the series
\SR{}, since, among the three \textit{A}Fe$_{2}$As$_{2}$ stoichiometric
compounds, $A=\mathrm{Rb}$ is the one that displays the strongest
evidence for $B_{1g}$ nematic order,\cite{Liu2018,Shibauchi18} as
compared to $A=\mathrm{Cs}$ and $A=\mathrm{K}$. The choice of Sr
is to minimize steric effects related to the different sizes of the
cations. To perform our DFT calculations, we used for the stoichiometric
compounds \Rb{} and \Sr{} the experimental room-temperature structures,
both belonging to the \textit{I4/mmm} space group (Rb: $a=3.87198\,\mathrm{\AA}$,
$c=14.46\,\mathrm{\AA}$, $z_{\mathrm{As}}=-0.1525$; Sr: $a=3.9243\,\mathrm{\AA}$,
$c=12.3644\,\mathrm{\AA}$, $z_{\mathrm{As}}=-0.14$).\cite{Tegel2008,Eilers2016}
The structural parameters of the series \SR{} were obtained by linear
interpolation of the lattice parameters and all bond lengths following
the equations: 
\begin{equation}
\begin{array}{rl}
\hat{a}_{\mathrm{int}} & =\hat{a}_{1}+x\cdot(\hat{a}_{2}-\hat{a}_{1})\\[7pt]
\mathbf{r}_{\mathrm{int}} & =\mathbf{r}_{1}+x\cdot(\hat{a}_{\mathrm{int}})^{-1}\hat{a}_{2}(\mathbf{r}_{2}-\mathbf{r}_{1})
\end{array}\label{e:str_int}
\end{equation}

Here, $\hat{a}$ is the matrix of the lattice vectors and $\mathbf{r}$
is a vector of fractional coordinates of a given atom in the supercell.
The indices $1$, $2$ and ``int'' refer to \Sr{}, \Rb{} and
\SR{}, respectively. For \SR, the spacer cation is replaced by
an artificial atomic species with a fractional nuclear charge of $Z=38-x$,
which simulates a random mixture of Sr and Rb using the virtual crystal
approximation.

\begin{figure}[h]
\centering \includegraphics[width=0.9\columnwidth]{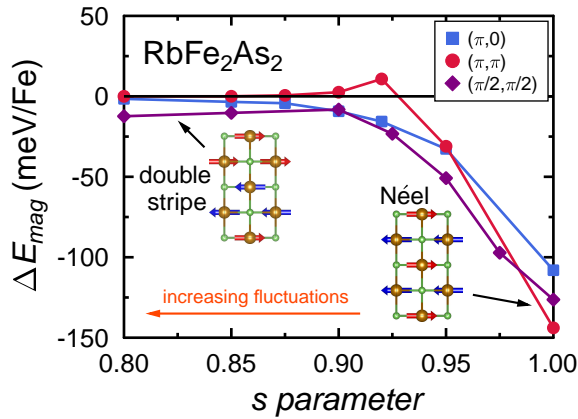} \vspace{-15pt}
 \caption{Total energies of the magnetic orders in \Rb{} with different ordering
vectors: $(\pi,0)$ (single stripe), $(\pi,\pi)$ (N\'eel order), and
$(\pi/2,\pi/2)$ (double stripe) relative to the non-magnetic state 
($\Delta E_\text{mag}=E_{\text{mag}}-E_{\text{non-mag}}$). 
The insets show the Fe spin configuration for the double-stripe and N\'eel orders. 
The energies are calculated for the experimental structure as functions of the
$s$ parameter of the reduced Stoner theory (RST). The GGA limit corresponds
to $s=1$, while $s<1$ leads to reduced Fe moments.}
\label{f:magnetic_energies_vs_r} 
\end{figure}

The competition between the non-magnetic state and three different
types of magnetic states (single-stripe, Néel, and double-stripe)
in this compound series was studied using the FLAPW method as available
in the \textsc{Elk} code.\cite{elk} Fully relativistic magnetic energies
were calculated using the PBE parameterization\cite{Perdew1996} of
the GGA density functional. The number of empty states was set to
24 per atom and spin, whereas the smearing was set to 0.002\,Ha$\,\approx\,$50\,meV
and the \textit{k}-mesh dimensions to $(10\times10\times10)$. The
aforementioned magnetic configurations \textendash{} single-stripe,
Néel, and double-stripe \textendash{} are the usual competitors in
Fe-based superconductors and have different spatial periodicities
corresponding to the ordering vectors (in the coordinate system of
the Fe-only square lattice) $\mathbf{Q}_{1}=(\pi,0)$, $\mathbf{Q}_{2}=(\pi,\pi)$
and $\mathbf{Q}_{3}=(\pi/2,\pi/2)$, respectively. In this work, these
states are simulated using the minimal 8-Fe supercells of different
shapes (see insets in Figs.~\ref{f:phase_diagram} and \ref{f:magnetic_energies_vs_r}). 
Each supercell includes a second Fe layer where all spins point 
in the opposite directions, which is commonly observed in iron pnictides.

One important ingredient of the physics in the iron-based superconductors
that is often not captured by standard DFT approaches is the role
of spin fluctuations. They are essential for instance to explain the
reduced magnetic moment of the Fe atoms, which is usually overstimated
in DFT.\cite{Ortenzi2015} Here, spin fluctuations are modeled using
the reduced Stoner theory\cite{Ortenzi2012} with a single adjustable
parameter $s$ that scales down the spin splitting of the self-consistent
potential $V(\mathbf{r})$ which enters then the DFT equations as:
\begin{equation}
V'_{\uparrow}(\mathbf{r})-V'_{\downarrow}(\mathbf{r})=s(V_{\uparrow}(\mathbf{r})-V_{\downarrow}(\mathbf{r}))\label{e:reduced_Stoner_theory}
\end{equation}

The role of $s$ is essentially to enhance the amplitude of the magnetic
fluctuations, which in turn results in a suppression of the Fe magnetic
moments. For the family of doped CaFe$_{2}$As$_{2}$ compounds, it
has been found that values around $s=0.85$ provide a consistent and
precise description of the Fe magnetic moments as a function of temperature
and doping.\cite{Ortenzi2015} For this reason, hereafter we set $s=0.85$
in our calculations. It turns out that correctly capturing these enhanced
spin fluctuations is crucial for the magnetic ground state, as indicated
by the calculated magnetic energies for \Rb{} versus the $s$ parameter
(Fig.~\ref{f:magnetic_energies_vs_r}). For the values around $s=0.85$,
the double-stripe is the dominating magnetic configuration showing
the correct order of magnitude of Fe moments $\sim\,0.5\,\mu_{\mathrm{B}}$.\footnote{We note that, in our simulations, the single-stripe order remains
the ground state of \Sr{} even for smaller values of $s$.} 
At the same time, the single-stripe and N\'eel orders strongly compete
with this double-stripe ground state. In contrast, $s=1$ leads to
overestimated Fe moments of the order of $2\mu_{B}$, commonly observed
in previous DFT simulations of iron-based systems,\cite{Opahle2009}
and the nature of the magnetic ground state of \Rb{} changes dramatically.
Nevertheless, our results are robust for a reasonably wide range of
values of $s$ corresponding to small Fe moments usually measured
in experiments.\cite{Cruz2008}

\begin{figure}[h]
\centering \includegraphics[width=0.9\columnwidth]{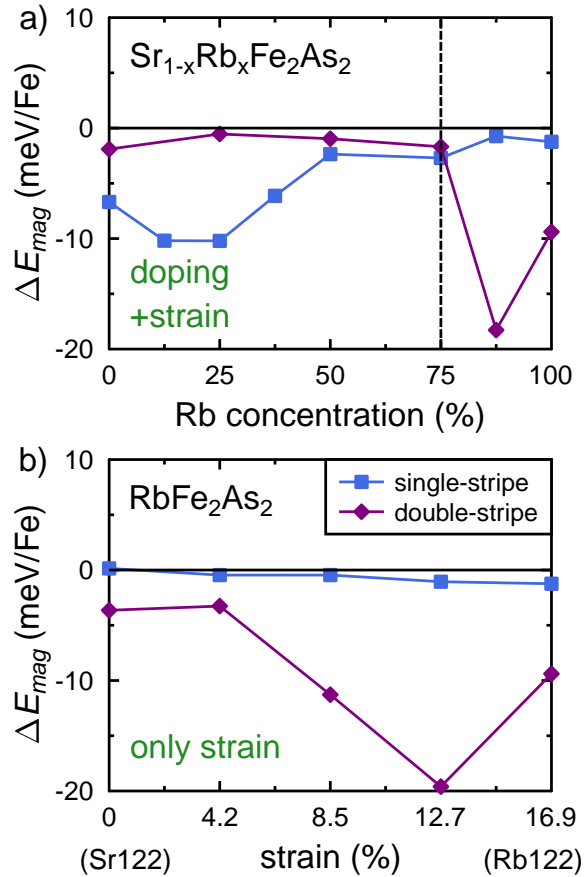} \vspace{-15pt}
 \caption{Evolution of magnetic energies relative to the non-magnetic state
($\Delta E_\text{mag}=E_{\text{mag}}-E_{\text{non-mag}}$) for the single- (squares)
and double-stripe (diamonds) orders in (a) the series \SR{} and
(b) \Rb{} under $c$-axis strain. In both cases, the structure is
tuned according to Eq.\,(\ref{e:str_int}). The parameter in the
reduced Stoner model, Eq.\,(\ref{e:reduced_Stoner_theory}), is fixed
to $s=0.85$, which brings the Fe moments close to the typically measured
values below $0.7\,\mu_{\mathrm{B}}$.}
\label{f:magnetic_energies_vs_x} 
\end{figure}

In the next step, we study the evolution of magnetic energies for
\SR{} with simultaneous hole-doping $x$ and lattice deformation
(Fig.~\ref{f:magnetic_energies_vs_x}a). As shown in Fig.~\ref{f:magnetic_energies_vs_x}a,
the single-stripe order of the parent \Sr{} compound is the ground
state up to a Rb concentration of about 75\%, where the magnetic ground
state switches to the double-stripe. At this point, the Fe occupation
is $n_{d}=5.625$, compared to $n_{d}=6$ for pure \Sr{}. We emphasize
here again that the N\'eel state is a relatively close competitor of
the double-stripe phase in the case of \Rb{} (Fig.~\ref{f:magnetic_energies_vs_r}).

In these simulations the effects of the electron count and the lattice
parameters are both present. In order to disentangle them and determine
which one plays the dominant role in driving the instability towards
the double-stripe order in \Rb{}, we compute the magnetic energies
when the lattice parameters are tuned between those of \Sr{} and
\Rb{}, according to Eq. (\ref{e:str_int}), and the Fe electron
count is fixed at $n_{d}=5.5$. The main change in the lattice parameters
is a stretch of the \textit{c}-axis corresponding to a strain of ${\sim}17\%$.
Importantly, such a stretch also affects various bond lengths. As
shown in Fig.~\ref{f:magnetic_energies_vs_x}b, for all strain values
investigated, the energy of the single-stripe magnetic state is always
higher than that of the double-stripe state. While the energy difference
is visibly affected by the $c$-axis lattice parameter, it is clear
that the main contribution to the stabilization of the double-stripe
order in \Rb{} arises from the change in the electron count. This
is further illustrated in Fig.~\ref{f:phase_diagram}, where both
parameters \textendash{} electron count $n_{d}$ and strain \textendash{}
are varied independently. While we observe that strain alone can change
the nature of the magnetic ground state for a fixed (smaller) hole-doping
level, there is a clear trend that large hole-doping favors a double-stripe
order. one. These results indicate that while bond lengths play a
significant role in choosing the magnetic ground state of \Rb{},
the electron filling of the Fe 3\textit{d} orbitals appears to be
the dominant factor determining the type of magnetic order.

\emph{Vestigial nematic orders. }So far our DFT simulations show a
change in the magnetic ground state of \SR{} from single-stripe,
for small and moderate $x$, to double-stripe for large $x$. As we
now argue, this evolution of the magnetic phase naturally accounts
for the change in the nematic state from $B_{2g}$ (small and mo\-derate
$x$) to $B_{1g}$ (large $x$) observed experimentally. The key point
is that both the single-stripe and the double-stripe states support
symmetry-breaking composite order parameters that can condense even
in the paramagnetic phase, forming so-called vestigial phases.\cite{Fernandes18}

The case of the single-stripe phase has been widely studied \cite{Kivelson08,Sachdev08,Fernandes12,Dagotto13}.
The Ising-like composite spin order parameter $\Phi_{B_{2g}}=\left\langle \mathbf{S}_{\mathbf{R}_{i}}\cdot\mathbf{S}_{\mathbf{R}_{i}+\mathbf{x}}\right\rangle -\left\langle \mathbf{S}_{\mathbf{R}_{i}}\cdot\mathbf{S}_{\mathbf{R}_{i}+\mathbf{y}}\right\rangle $
is clearly non-zero and uniform in the single-stripe phase, since
spins are parallel to each other along one axis and anti-parallel
to each other along the other axis (see Fig. \ref{fig_vestigial}a).
As shown by a variety of different theoretical methods \cite{Kivelson08,Sachdev08,Fernandes12,Dagotto13},
$\Phi_{B_{2g}}$ can remain non-zero even when there is no long-range
magnetic order, $\left\langle \mathbf{S}_{\mathbf{R}_{i}}\right\rangle =0$,
because $\Phi_{B_{2g}}$ is a two-spin correlation function. Now,
since $\Phi_{B_{2g}}$ breaks the equivalence between the $x$ and
$y$ axis, its onset breaks tetragonal symmetry, while preserving
translational symmetry. The state with $\Phi_{B_{2g}}\neq0$ but $\left\langle \mathbf{S}_{\mathbf{R}_{i}}\right\rangle =0$
is a $B_{2g}$ nematic state. It is called a vestigial phase of the
magnetically ordered state because it breaks a subset of the symmetries
broken by the latter, as explained in the Introduction.

Analogously, inside the double-stripe magnetic phase, the Ising-like
composite order parameter $\Phi_{B_{1g}}=\left\langle \mathbf{S}_{\mathbf{R}_{i}}\cdot\mathbf{S}_{\mathbf{R}_{i}+\mathbf{x}+\mathbf{y}}\right\rangle -\left\langle \mathbf{S}_{\mathbf{R}_{i}}\cdot\mathbf{S}_{\mathbf{R}_{i}-\mathbf{x}+\mathbf{y}}\right\rangle $
is non-zero and uniform \cite{Paul11}, since spins are parallel along
one diagonal and anti-parallel along the other diagonal (see Fig.
\ref{fig_vestigial}b). In this case, a vestigial $B_{1g}$ nematic
phase, in which diagonals are not equivalent, can appear even in the
absence of long-range magnetic order~\cite{Fernandes17,Dagotto17}.
Thus, the change in the magnetic ground state from single-stripe to
double-stripe should be accompanied by a change in the vestigial nematic
order from $B_{2g}$ to $B_{1g}$.

An interesting question that remains to be addressed is why \Rb{}
shows nematic order but no apparent sign of long-range magnetic order,
whereas \Sr{} displays both magnetic and nematic orders.\cite{Wu2014}
The situation seems to be similar to the case of the Ti-based oxypnictides
\cite{Fernandes17}, where a $B_{1g}$ vestigial nematic phase driven
by double-stripe magnetic fluctuations has been proposed, but long-range
magnetic order has not been observed. First, it is important to recall
that the nematic transition, which must always happen above or simultaneously
to the magnetic transition, is observed at lower temperatures in \Rb{}
($T_{\mathrm{nem}}\sim40$\,K)\cite{Liu2018} as compared to \Sr{}
($T_{\mathrm{nem}}\sim200$\,K).\cite{Wu2014} Second, the existence
of a closely competing N\'eel state in the case of \Rb{} (Fig.~\ref{f:magnetic_energies_vs_r})
provides a possible mechanism for the suppression of the magnetic
transition temperature, similarly to what has been proposed to explain
the absence of magnetic order in FeSe.\cite{Glasbrenner15}

Finally, the experimental\cite{Eilers2016,Meingast16} and theoretical
indications\cite{Hardy2013,Medici2014,Backes2015} that electronic
correlations are stronger in hole-doped \AFeAs{} iron pnictides
as compared to \AeFeAs{} could also be an aspect contributing to
the change in nature of nematicity in these systems. In the present
work, we concentrated on the underlying magnetism and found strong
hints of its role in determining the various nematic phases.

\emph{Conclusions. }In summary, our first-principles calculations
of the magnetic ground states of 122-type iron-based compounds show
that the dominant magnetic instability in these systems is strongly
affected by the nature of the spacer cation substitution. For \Rb{},
we found that the double-stripe state becomes the leading instability,
in contrast to the single-stripe order in \Sr{}. Analysis of the
various contributions involved in the calculations showed that the
magnetic order is more sensitive to the electron count variation due
to the Sr$\to$Rb substitution than to the related structural changes.
Based on these results, we proposed that the evolution of the nematic
state from $B_{2g}$ for \Sr{} to $B_{1g}$ for \Rb{} is a consequence
of the change in the vestigial ordered states supported by the single-stripe
and double-stripe orders.

Our results have important experimental consequences that can in principle
be verified. First, even though \Rb{} is not magnetic, it should
have strong magnetic fluctuations peaked at the $\mathbf{Q}_{3}=\left(\pi/2,\pi/2\right)$
wave-vector. To the best of our knowledge, there are no inelastic
neutron scattering measurements available for \Rb{}. Neutron scattering
in the related $n_{d}=5.5$ KFe$_{2}$As$_{2}$ compound, which shows
no evidence of $B_{1g}$ nematic order, reveal incommensurate spin
fluctuations peaked at $(\pi(1\pm2\delta),0)$ with $\delta\approx0.16$.\cite{Lee2011}
Also, our phase diagram in Fig. \ref{f:phase_diagram} shows that
uniaxial strain along the $c$-axis can tune the magnetic ground state
for compounds with intermediate levels of hole doping. This offers
the possibility to tune the nematic state from $B_{2g}$ to $B_{1g}$
as a function of strain or pressure.

\emph{Acknowledgments.} We thank T. Shibauchi and J.~Schmalian for fruitful discussions. 
VB and RV were financially supported by DFG Sonderforschungsbereich TRR 49. 
The computer time was allotted by the centre for supercomputing (CSC) 
in Frankfurt and by the computer center of Goethe University. 
RMF was supported by the U.S. Department of Energy, Office of Science, 
Basic Energy Sciences under award number {DE-SC0012336}. Parts of some figures 
have been produced with \textsc{VESTA3} (Ref.~\onlinecite{Vesta}).

\vfill{}

\bibliographystyle{prb-titles}
\bibliography{paper}

\end{document}